\documentstyle[aps,twocolumn]{revtex}
\begin{document}
\title{Universal Thermodynamics of Degenerate Quantum Gases in the Unitarity Limit}
\author{Tin-Lun Ho}
\address{Department of Physics,  The Ohio State University,
Columbus, Ohio
43210}

\maketitle

\begin{abstract}
We perform a systematic study of the thermodynamics of quantum gases in the unitarity limit. 
Our study makes use of a "Universality Hypothesis" for the relevant energy scales of a many-body system at unitarity. This Hypothesis is supported by recent experiments and can be proven in Boltzmann regime. It implies a universal  form for the grand potential which is characterized by only a few universal numbers in degenerate limit.  This universal form  
provides a simple way to determine the density profile of a trapped fermion superfluid
as well as the second sound velocity of a homogeneous superfluid at unitarity. 
\end{abstract}

Feshbach resonance has introduced a whole new  dimension in the 
research of degenerate quantum gases.  Through this resonance, effective interactions between atoms are dramatically increased. 
Such resonance arises when the energy of a pair of scattering atoms is tuned close to that of a molecular bound state by an external magnetic field, thus  causing substantial resonance scattering.  At resonance,  
scattering reaches the {\em unitarity limit}:  with  a diverging scattering length $a_{sc}$, and a cross section reaching the maxium value $4\pi/k^2$, where $k$ is the relative momentum of the scattering atoms. These properties are {\em universal} because they are independent of any feature of atomic potentials. 

This universality, simple as it is, poses a challenging many-body problem,  as 
there are no small perturbative parameters. On the other hand, it can lead to great simplification if one makes a very reasonable assumption (referred to as 
{\em ``Universality Hypothesis"}) : that the only dominant length scale at unitarity in the ground state is inter-particle spacing $n^{-1/3}$, where $n$ is the density.  
This is based on the idea that the only relevant length scales in the ground state are $a_{sc}$ and  $n^{-1/3}$.   Since $a_{sc}$ diverges at resonance, it must disappear from all physical quantities, leaving $n^{-1/3}$ the only relevant length scale.  
The word ``Hypothesis" is to indicate that although universality has emerged in approximate calculations\cite{Holland}\cite{Heiselberg}, it has not been proven rigorously except in Boltzmann regime\cite{HoMueller}.  This hypothesis also implies that for both bosons and fermions, the only relevant energy scale is 
the ``Fermi" energy ${\cal E}_{F}^{o}(n) \equiv (\hbar^2/2M)(3\pi^2 n)^{2/3}$.  For the same reason, the transition temperature $T_{c}$ of a fermi superfluid must scale as $T_{F}= {\cal E}_{F}/k_{B}$, i.e.  $T_{C} =\gamma T_{F}$,  where $\gamma$ is a universal constant. 
That $ \gamma$ can be of order 1 was suggested by Holland et.al\cite{Holland}. Current estimates of $T_{c}$ range from 0.5 to 0.2$T_{F}$\cite{Holland}\cite{Griffin}. The possibility of such a high  $T_{c}$ has made Feshbach resonance a focus of attention in the current race for achieving fermion superfluidity.  On the other hand, the normal state is no less intriguing, 
for it contains the same non-perturbative effects. 

The universal properties of a Fermi gas in unitarity regime have been demonstrated recently by  a sequence of beautiful experiments\cite{Duke}\cite{ENS}\cite{MIT}\cite{Jin1}. John Thomas's group\cite{Duke} has pointed out that that the interaction energy $E_{int}$ of a Fermi gas of $^{6}$Li near Feshbach resonance is the form $\beta{\cal E}_{F}^{o}$, $\beta\approx -0.25$  over the temperature range $1>T/T_{F}>0.1$.  A direct measurement of this energy was performed later by Salomon's group  who found a similar energy at higher temperatures which remains {\em smooth} across the resonance. 
Not only does $E_{int}$ saturate at resonance despite the diverging scattering length, it scales directly with the Fermi energy. Similar saturation are also observed in RF spectroscopy by Ketterle's group\cite{MIT} and Jin's group\cite{Jin1}.
The observed relation between  $E_{int}$ and ${\cal E}_{F}^{o}$ 
 is a support for the Universality Hypothesis.  Further experiments on other alkali fermions will help to verify its validity.
The sign of $E_{int}$ and its continuity across the resonance\cite{ENS}, however, 
require additional physics and are related to formation of molecules\cite{HoMueller}.  
Since the Duke experiments cover the temperature range above and below the estimated $T_{c}$, it raises the question of how superfluidity  is affected by unitarity, and their 
signature in the unitarity regime.

In this paper, we perform a systematic study of the thermodynamics of quantum gases at unitarity  using the Universality Hypothesis.  
%While many relations below are applicable to both bosons and fermions,  we mostly focus on %Fermi gases. 
We shall show that: {\bf (A)} At unitarity, the thermodynamic potentials acquire universal forms which depend only on the nature of the thermodynamic phase.  {\bf (B)} The properties of a degenerate Fermi gas near resonance (be it normal or superfluid) are characterized by only a few universal numbers.
{\bf (C)}  Universal thermodynamics provides a simple way to determine the density profile of a trapped fermion superfluid near resonance.  {\bf (D)}  It also allows one to calculate the hydrodynamic modes of a fermion superfluid at finite temperature.  {\bf (E)} Bose gas 
in the unitarity limit, if stable, will have a {\em fermionic} energy density. 

Before proceeding, we first make clear what quantities the 
Universality Hypothesis describes. Let us consider the Hemholtz free energy density $f=f(T, n, B, \{ r_{i} \})$, which is a function of $T$, $n$, external magnetic field (which controls $a_{sc}$), and other interaction lengths such as effective range, etc, 
(collectively denoted as $\{ r_{i} \}$).  While $a_{sc}$ diverges at resonance (say, at  $B=B_{o}$), it is assumed that  $\{ r_{i} \}$ remain atomic size so that   
$x_{i}=r_{i}^{}n^{1/3}<< 1$.  If $f$ is smooth across the resonance (as indicated in ref.\cite{ENS}), then in the neighborhood of $B_{o}$ it is well approximated by 
its value at resonance $f(T,\mu; B_{o}, \{ x_{i} \})$.  Moreover, if $f$ has an 
asymptotic expansion in $x_{i}$, we then have 
\begin{equation}
f(T,n; B_{o}, \{ x_{i} \}) =f(T,n; B_{o}, 0) \left( 1 + 0(x_{i}) \right).
\label{1} \end{equation}
Universality Hypothesis describes the first term eq.(\ref{1}) $--$ that it has only two energy scales,  ${\cal E}_{F}^{o}(n)$ and $k_{B}T$; and is independent of $B_{o}$ for it would otherwise introduce additional energy scales. In this paper, quantities of ideal gas will be denoted by a superscript $``^{o}"$. 

{\bf I. Universal Thermodynamics:}  
We begin by deriving some general thermodynamic properties 
%In this section, we derive some general properties of the thermodynamics 
at resonance valid for different phases. To be efficiency, we consider the grand potential instead of the Hemholtz free energy. The former is defined as 
 $\Omega$$(T,\mu_{\uparrow}, \mu_{\downarrow},V)$$=$$- k_{B}T$${\rm ln}$${\rm Tr}$
$e^{-(H- \mu_{\uparrow}N_{\uparrow} - \mu_{\downarrow}N_{\downarrow} )/k_{B}T}$ for a two component quantum gas (labelled as $\uparrow$ and $\downarrow$) with identical mass $M$ in a volume $V$.  We first consider $\mu_{\uparrow}=\mu_{\downarrow}$ which implies
 $n_{\uparrow}=n_{\downarrow}=n/2$, or $m=(n_{\uparrow}-n_{\downarrow})/2=0$.   
According to Universality Hypothesis, all 
microscopic scales are absent at resonance. The only energy scales are $k_{B}T$ and $\mu$. The corresponding density scales are $\lambda^{-3}$ and $n_{f}(\mu)$, where $\lambda= h/\sqrt{2\pi M k_{B}T}$ is the thermal wavelength, and $n_{f}(\mu) \equiv (3\pi^2)^{-1}(2M  \mu/\hbar)^{3/2}$. Since pressure $P=-\Omega/V$, we have the following two equivalent relations from dimensional analysis
\begin{equation}
P(T,\mu) = \frac{2 k_{B}T}{5 \lambda^3} {\cal W}_{0}(x^{-1}) =  \frac{2\mu n_{f}(\mu)}{5}{\cal G}_{0}(x), 
\label{P}\end{equation}
where $x=  k_{B}T/\mu$,  and (${\cal W}_{0}$, ${\cal G}_{0}$) are dimensionless scaling function.  These two forms are useful in 
Boltzmann and degenerate regime respectively since their arguments are small in these cases. 
Using the well known relation $\epsilon = Ts + \mu n  -P$  for energy density $\epsilon$
and the Gibbs-Duham relation ${\rm d}P = n{\rm d}\mu + s{\rm d}T$,  we have 
\begin{equation} 
n = n_{f}(\mu) \left( {\cal G}_{0}(x) - \frac{2}{5} x{\cal G}_{0}'(x) \right) 
\label{n} \end{equation}
 \begin{equation}
 s =  \frac{2n_{f}(\mu) k_{B}}{5} {\cal G}_{0}'(x),  \,\,\,\,\,\,\,\,\,\, \epsilon = \frac{3}{2} P.
\label{epsilon} \end{equation}
Eq.(\ref{P}) to (\ref{epsilon})  are identical to those of idea gas, where   $k_{B}T$ and $\mu$ 
are also the only energy scales.  However, unlike ideal gas, the absence of microscopic energy scale here is not due to absence of interaction, but instead the  maximum scattering between particles.  Since universality hypothesis makes no reference to the thermodynamic phase, eq.(\ref{P}) to (\ref{epsilon})  {\em apply to both normal and superfluid phases}, which  have of course, different scaling functions. The scaling functions, however, are constraint by the positivity of $s$ and $n$, as well as stability conditions $\partial^2 P/\partial T^2 = \partial s/\partial T >0$ and  
$\partial^2 P/\partial \mu^2 =\partial n/\partial \mu >0$. 

The density profile $n({\bf x})$ in a non-uniform potential $V({\bf x})$ can be readily determined from eq.(\ref{n}) within local density approximation (LDA) by replacing  $\mu$ with $\mu({\bf x})\equiv \mu - V({\bf  x})$.  The total energy of a system of $N$ particles 
$E=E(T,N)$ can the be obtained by eliminating $\mu$ from the relations 
$N=\int n({\bf x})=N=N(T, \mu)$ and   $E=\int \epsilon({\bf x}) = \int  3P({\bf x})/2
=E(T, \mu)$. 

When $m \neq 0$,  Universality Hypothesis implies that 
\begin{equation}
P(T, \mu_{\uparrow}, \mu_{\downarrow}) =\frac{2\mu n_{f}^{}(\mu)}{5} {\cal G}(x; \nu/\mu) 
\label{Pmunu}\end{equation}
where $\mu= (\mu_{\uparrow}+\mu_{\downarrow})/2$, $\nu= \mu_{\uparrow}-\mu_{\downarrow}$, and $ {\cal G}(x; \nu/\mu)$ is a scaling function even in $\nu$ due to the invariance of $\Omega$ under spin exchange.   For small $\nu$, we have
\begin{equation}
{\cal G}(x; \nu/\mu) = 
{\cal G}_{0}(x) + \frac{{\cal G}_{2}(x)}{2} \left(\frac{\nu}{\mu}\right)^2 + ... 
\label{nu2} \end{equation}
where ${\cal G}_{2}$ is another dimensionless function. Defining magnetic  susceptibility $\chi$ and specific heat $c$ at constant $\mu$ as $m= (n_{\uparrow}- n_{\downarrow})/2=\partial P/\partial \nu = \chi \nu$ and  $c = T(\partial s/\partial T)_{\mu}$, eq.(\ref{Pmunu}) and (\ref{nu2}) then implies
$c/ (T\chi) =k_{B}^2 {\cal G}''_{0}(x)/{\cal G}_{2}(x) $, 
which is a universal function of $x= k_{B}T/\mu$.  

{\bf II. Boltzmann limit:}  This is the limit where the fugacities $z_{i}= e^{\mu_{i}/k_{B}T}$, $i=\uparrow, \downarrow$ are small; and where universality can be proved rigorously\cite{HoMueller}.  We shall first generalize the derivation in ref.\cite{HoMueller} to arbitrary spin polarization and the derive the key thermodynamic properties in the Boltzmann regime. 
Expanding $\Omega$ in $z_{i}$\cite{LL} for a Fermi gas, we have  
\begin{equation}
P(T, \mu_{\uparrow}, \mu_{\downarrow}) = P^{(o)}(T, \mu_{\uparrow}, \mu_{\downarrow}) + 2\sqrt{2} b_{2} \frac{k_{B}T }{\lambda^3} z_{\uparrow} z_{\downarrow} 
\label{PBolt} \end{equation}
where $P^{(o)}(T, \mu_{\uparrow}, \mu_{\downarrow})= \sum_{i=\uparrow, \downarrow}
k_{B}T \lambda^{-3}\left( z_{i} - 2^{-5/2} z_{i}^2\right)+ O(z_{i}^3)$ is the pressure of the an ideal Fermi gas, and $b_{2}$ is the second virial coefficient which is a function of temperature only. 
Using the relations $\epsilon= Ts + \mu_{\uparrow}n_{\uparrow} + \mu_{\downarrow}n_{\downarrow}-P$ and  ${\rm d}P = \sum_{i} n_{i}{\rm d}\mu_{i} + s{\rm d}T$, we have 
\begin{equation}
n_{\uparrow(\downarrow)} \lambda^3 = z_{\uparrow(\downarrow)}
\left( 1 + 2\sqrt{2} b_{2}z_{\downarrow(\uparrow)}\right) - 2^{-3/2} 
z_{\uparrow(\downarrow)}^2, 
\label{xxx} \end{equation}
\begin{equation}
s= \frac{5}{2}\frac{P}{T} - \frac{\mu_{\uparrow}n_{\uparrow} + \mu_{\downarrow}n_{\downarrow}}{T} +2\sqrt{2}  \frac{k_{B}T}{\lambda^3}\frac{\partial b_{2}}{\partial T}z_{\uparrow}z_{\downarrow} , 
\end{equation}
\begin{equation}
 \epsilon= \frac{3P}{2}+ 2\sqrt{2}  \frac{k_{B}T}{\lambda^3} z_{\uparrow}z_{\downarrow}
T \frac{\partial b_{2}}{\partial T}. 
\label{epxxx} \end{equation}
Since $b_{2}=1/2$, and $\partial b_{2}/\partial T=0$ at resonance\cite{LL}\cite{HoMueller}, we recover the universal thermodynamics in  {\bf I}. 

It is also useful to use $(T,n)$ instead of $(T, \mu)$ as variables. The following relations are applicable to all scattering lengths and can be compared with experiments. 
Eliminating $z_{i}$ in eq.(\ref{xxx}) to (\ref{epxxx}), we have 
\begin{equation}
P=k_{B}T( n  + [  2^{-5/2}(n_{\uparrow}^2 + n_{\downarrow}^2) 
- 2\sqrt{2} b_{2}  n_{\uparrow}n_{\downarrow}] \lambda^3), 
\label{eqos}\end{equation}
\begin{equation}
\epsilon= 
\frac{3}{2}k_{B}T\left[ n + \frac{(n_{\uparrow}^2 + n_{\downarrow}^2)\lambda^3}{2^{5/2}} - \phi n_{\uparrow}n_{\downarrow}\right], 
\label{epBolt} \end{equation}
where $\phi =  2\sqrt{2} \lambda^3(  b_{2} - \frac{2}{3} 
T\partial b_{2}/\partial T) $. 
From eq.(\ref{eqos}) and (\ref{xxx}), it is easy to calculate isothermal compressibility $\kappa_{T}^{}$
$=$$n^{-1}(\partial n/\partial P)_{T}$ and  isothermal spin susceptibility $\chi_{T}^{}$$ =$ 
$(\partial m/\partial \nu)_{T}$. Their deviations from ideal gas values are 
$\Delta\kappa_{T}^{}$$=\sqrt{2}b_{2}n\lambda^3/(nk_B T)$, $\Delta \chi_{T}^{}$$= - (n/k_{B}T)$
$(\sqrt{2} b_{2} n\lambda^3/4)$, hence $n^2 \Delta\kappa^{}_{T} /\Delta\chi^{}_{T}=-4$.
 It is easy to derive the same results for the Bose gas, which is eq.(\ref{eqos})
and (\ref{epBolt}) with a minus sign in the $n_{i}^2$ terms. 

{\bf III.  Degenerate Normal Gas:}  This is the case  $x= k_{B}T/\mu<<1$.  For small spin polarization, the pressure can be expanded in $x$ and $\nu/\mu$ as $P=P^{(n)}(T, \mu, \nu)$,
 \begin{equation}
P^{(n)} = \frac{2\mu n^{}_{f}(\mu)}{5}A^{3/2}\left[1 + \frac{5\pi^2 (Bk_{B}T)^2}{8(A\mu)^2}  +  \frac{15 (C\nu)^2}{32(A\mu)^2} + ..\right].
\label{Pn}\end{equation}
where coefficients $A$, $B$, $C$ are universal numbers. They are written in this form to simplify later discussions. The absence of linear $T$ term is due to vanishing entropy at $T=0$. For ideal Fermi gas, $A=B=C=1$.  Stability conditions $\partial s/\partial T$, $\partial n/\partial \mu>0$ imply that   $A,B> 0$; and $C>0$ unless the system is ferromagnetic.  
From  eq.(\ref{n})  and the relation $m= \partial P/\partial \nu$, we have  
$n = n^{(n)}(T, \mu, \nu)$, 
\begin{equation}
n^{(n)} =n^{}_{f}(A\mu) \left[ 1 +  \frac{\pi^2 (Bk_{B}T)^2}{8(A\mu)^2}  +  \frac{3}{32} 
\left(\frac{C\nu}{A\mu}\right)^2 +.. \right], 
\label{nTmu} \end{equation}
and $m=[3c/(8A)] [n^{}_{f}(A\mu)/(A\mu)](C\nu)$,  where $n_{f}^{}(A\mu)=A^{3/2}n_{f}^{}(\mu)$. 
These two equations for $n$ and $m$ readily give the density profile 
$n_{i}({\bf x})$ ($i = \uparrow, \downarrow$) in non-uniform potentials $V_{i}({\bf x})$ within LDA 
by substituting $\mu_{i}({\bf x})= \mu_{i} - V_{i}({\bf x})$. Note, however, 
that these relations are valid only for $\mu_{i}({\bf x}) >>k_{B}T$ and $\nu<<\mu$.   
As one approaches the surface of the cloud, 
density decreases and the system will switch to Boltzmann regime in surface regions where
$e^{\mu_{i}({\bf x})/k_{B}T}<<1$, with densities given by eq.(\ref{xxx}). 

To find an accurate formula interpolating between degenerate and Boltzmann limit, we note that (considering the case $\nu=0$ for simplicity) deep in Boltzmann regime,  $z=e^{\mu/k_{B}T}<<1$,  eq.(\ref{xxx}) is simply $n =z/\lambda^3$ and is the high temperature limit of the ideal gas relation
$n= n_{id}^{}(\mu, T)$, $n_{id}(\mu, T) = \lambda^{-3}f_{3/2}^{}(z)$, where $f_{3/2}(z)$ is the  Fermi integral\cite{Fermif}.  On the other hand, eq.(\ref{nTmu}) is precisely the low temperature expansion of the ideal gas relation $n_{id}(\mu, T)$ with the substitution $(\mu, T)\to ( A\mu, BT)$. 
The desired interpolation will then be of the form 
\begin{equation}
n^{(n)}(T,\mu) = n_{id}^{}(A(x)\mu, B(x)T),  \,\,\,\,\,\, x=\mu/k_{B}T
\label{nint} \end{equation}
where $(A(x), B(x))$ are functions of $x$ (as required by Universality Hypothesis) such that 
$(A(x), B(x))\to (A,B)$ as $x>>1$, and $(A(x), B(x))\to (1,1)$ as $z=e^{x}<<1$. Since the switching from degenerate to Boltzmann regime take place at $\mu\sim k_{B}T$, a simple  
extrapolation is 
\begin{equation}
A(x) = \frac{Ae^{x}+1}{e^{x}+1}, \,\,\,\,\,\,\, B(x) = \frac{Be^{x}+1}{e^{x}+1}.
\label{ABint}
\end{equation}
Eq.(\ref{nint}) and (\ref{ABint}) form the desired extrapolation.  The density profile in a trap calculated within LDA using eq.(\ref{nint}) and (\ref{ABint}) is shown as the dashed curve in  
the lower figure in figure 1. 

To derive relations related to experiments, we invert the relations $n=n(T, \mu, \nu)$,  
$m=m(T, \mu, \nu)$ to express $\mu$, $\nu$, and hence $\epsilon= 3P/2$  in terms of $(T,n,m)$.  To the lowest order in $k_{B}T/{\cal E}_{F}^{o}$, we have 
$\mu=\mu^{(n)}(T,\mu,\nu)$, 
\begin{equation}
\mu^{(n)}  = {\cal E}^{o}_{F}(n)(1- W)/A, \,\,\,\,\,\, \frac{C\nu}{A\mu} = \frac{8A}{3C} \frac{m}{n}
\end{equation}
\begin{equation}
\epsilon^{(n)}  =[ 3n {\cal E}^{o}_{F}(n)/5] \left[ 1 + 5W\right]/A, 
\label{epep} \end{equation}
where $W= \frac{\pi^2}{12}\left( \frac{Bk_{B}T}{{\cal E}_{F}^{o}} \right)^2 + \left( \frac{2A m}{3Cn}
 \right)^2$.  
Alternatively, we can write  $\mu^{(n)}= {\cal E}_{F}^{o}(1+ \beta_{\mu}^{})$ and 
$\epsilon^{(n)}= (3n{\cal E}_{F}^{o}/5)(1+ \beta_{\epsilon}^{})$. The numbers $\beta_{\mu}^{}$ and 
$\beta_{\epsilon}^{}$ are the interaction parameters measured in ref.\cite{Duke} and ref.\cite{ENS}. We show here that they have opposite temperature and  spin polarization corrections, which  differ from each other by a factor of 5.

 {\bf IV.  Superfluid at resonance}:  
This is the case where universal thermodynamics proves very useful. 
We shall consider superfluids with zero spin polarization, (hence $\nu=0$). 
Within Ginzburg-Landau theory, 
the difference in grand potential between a superfluid with order parameter $\Psi({\bf r})= \langle\psi_{\uparrow} \psi_{\downarrow}\rangle$ and a normal fluid nrar 
superfluid transition is $\Omega^{}[\Psi] -  \Omega^{(n)} = 
\int {\rm d}{\bf r} {\cal \omega}[\Psi({\bf r})]$, 
\begin{equation} 
\omega[\Psi] =  K|\nabla \Psi|^2 - r_{2}|\Psi|^2 + r_{4}|\Psi|^4/2
\label{omega} \end{equation}
where $K, r_{4}>0$, and $r_{2}$ vanishes at transition. 
The equilibrium potential is $\Omega^{} = \Omega [\Psi_{o}]$, where $\Psi_{o}$ is the minimum of eq.(\ref{omega}). According to Universality Hypothesis,  $K$, $r_{2}$ and $r_{4}$ are functions of $k_{B}T$ and $\mu$ only. The condition for transition $r_{2}(T=T_{c},\mu) =0$ implies $T_{c} = T_{c}(\mu)$.  Using dimensional analysis and expanding $r_{2}$, $r_{4}$ and $K$ about $T_{c}$, we have $k_{B}T_{c}(\mu)=\gamma (A\mu)$, $K=\xi  \hbar^2/(2M)$; 
and to the lowest order of $1-T/T_{c}$, 
$r^{}_{2} = \alpha{}_{2}\mu ( 1 - T/T_{c})$, $r^{}_{4} = \alpha{}_{4}\mu/n_{f}(A\mu)$, 
where ($\gamma$, $\xi$, $ \alpha^{}_{2}$, $\alpha^{}_{4}$) are universal numbers
characterizing the superfluid state near $T_{c}$.  

When $T>T_{c}$, the system is normal with 
$\Psi_{o}=0$,  and 
\begin{equation}
P= P^{(n)}(T,\mu, 0), \,\,\,\,\,\,\,\,\, n=n^{(n)}(T,\mu, 0). 
\end{equation}
 For $T<T_{c}$, we have $|\Psi_{o}|^2= r_{2}/r_{4}$, and 
$P= P^{(n)}(T,\mu) + r_{2}^2/(2r_{4})$.  Explicitly, we have 
$|\Psi_{o}|^2 = \alpha n_{f}(A\mu) (1-T/T_{c})$,   $\alpha= \alpha^{}_2 /\alpha^{}_4$,  and 
$P=P^{(s)}(T, \mu)$, 
\begin{equation}
P^{(s)}= P^{(n)}(T, \mu, 0)+ \frac{2\mu n^{}_{f}(A\mu) D }{5}  \left( 1-\frac{x}{\gamma}\right)^2,  
\label{Ps} \end{equation}
where   $D = 5\alpha_{2}^2/(4\alpha^{}_{4})$,  and $x=k_{B}T/\mu$. Eq.(\ref{Ps}) then implies
$n= n^{(s)}(T, \mu)$, 
\begin{equation}
n^{(s)} = n^{(n)}(T, \mu, 0) + 
n^{}_{f}(A\mu) D \left( 1-\frac{x}{\tilde{\gamma}} \right)   \left( 1-\frac{x}{5\tilde{\gamma}} \right)  , 
\label{ns} \end{equation}
where $\tilde{\gamma}=\gamma A$, $x=k_{B}^{}T/\mu$.
Eq.(\ref{ns}) and eq.(\ref{nint}) provide a simple method to construct the density profile of a trapped superfluid within LDA: We first plot $\mu({\bf r}) = \mu - V({\bf r})$ and $\tilde{\gamma}\mu({\bf r})$ as a function of ${\bf r}$. (See fig.1) 
The regions where  $k_{B}T<\tilde{\gamma}  \mu({\bf r})$ and $\tilde{\gamma} \mu({\bf r})<k_{B}T$
correspond to superfluid (SF) and normal (N) region.  The latter is further separated into 
degenerate normal regime (DN), $\tilde{\gamma}\mu>>k_{B}T$,  and Boltzmann (B) regime,  $e^{\mu/k_{B}T}<<1$. 
% (There are no sharp boundaries between them of course). 
The densities in (SF) and (N) are given by eq.(\ref{ns}) and (\ref{nint})  respectively. 
The ``superfluid bump" in fig.1 is also obtained in  ref.\cite{Holland}. Here, we show that it is a necessary consequence of the Universality Hypothesis and display its 
general structure. 

To express quantities in terms of $T$ and $n$, we invert eq.(\ref{ns}) and then obtain 
$\mu= \mu^{(s)}(T,n)$, $\epsilon= \epsilon^{(s)}(T,n)$.  If superfluid transition takes place in degenerate regime, ${\cal \epsilon}  \equiv  (\pi  B k_{B}^{} T_{c}^{})^2/(8(A\mu)^2)= 
 (\pi k_{B}^{}\gamma^{})^2/8<<1$. In that case, it is simple to show that 
\begin{equation}
\mu^{(s)}  = \mu^{(n)}(T, n, 0)  - \zeta {\cal E}_{F}^{o} \left( 1- \frac{y}{\gamma}\right) 
\left( 1- \frac{y}{5\gamma}\right)  
\end{equation}
\begin{equation}
\epsilon^{(s)}  =\epsilon^{(n)} (T, n, 0) - \frac{3n{\cal E}_{F}^{o}\zeta }{5} 
\left( 1 - \frac{y^2}{\gamma^2}\right) 
\label{epsilonsf} \end{equation}
where   $y= k_{B}T/{\cal E}_{F}^{o}(n)$, 
$k^{}_{B}T_{C}^{}= \gamma A\mu = \gamma {\cal E}_{F}^{o}(n)$, 
$\zeta = 2D/(3A)=5\alpha^{2}_{2}/(6A\alpha^{}_{4})$.
Eq.(\ref{epsilonsf}) and (\ref{epep})  imply a universal jump $[(c^{}_{s} - c^{}_{n})/c^{}_{n}]_{T^{}_{c}} = 12\zeta A/(5\pi^2 \gamma^{2}B^{2})$
cross $T^{}_{c}$. 

In the superfluid phase,  a ``second" sound mode ($u_{2}$) must exist  in addition to the first (or ordinary) sound $u_{1}$.  In the collisional limit, their  velocities are 
$u_{1} = \sqrt{ ( \partial P/(M \partial n)_{\sigma} }$ and 
$u_{2} = \sqrt{ \sigma^2 \rho_{s}/(\rho_{n} (\partial \sigma/\partial T)_{P}) }$ 
respectively\cite{second}, where $\sigma= s/(Mn)$ is the entropy per unit mass, $\rho_{s}$ is  
the superfluid  mass density which can be obtained easily from the gradient  term of eq.(\ref{omega}) to be 
$\rho_{s} = \tilde{\xi}  M n ( 1- T/T_{c})$, $\tilde{\xi} = 4\xi  \sigma \alpha$\cite{rhos}; 
 $\rho_{n}$ is normal fluid mass density, 
$\rho_{s} + \rho_{n}= Mn$. We then have $u_{1}^{2}  =  2{\cal E}^{o}_{F} /(3MA)$, and  $u_{2}^{}/u_{1}^{}= 
Q \sqrt{1 - T/T_{c}}$, $Q^2 = \frac{3}{4}(\gamma B)^2 \tilde{\xi} \{1+ \frac{12\zeta}{5} 
\left( \frac{ A}{\pi \gamma B}\right)^2 \}^{-1}$
 \cite{commentcal}.

{\bf V. Bosons in unitarity limit:}  Since Universality Hypothesis makes no reference on statistics, 
eq.(\ref{Pn}) should also apply to {\em stable} Bose system at unitarity, ($\nu=0$ for single component gas).  However,  it  provides no information about phase coherence and hence does not guarantee that the system is a superfluid\cite{1D}. Nevertheless, it predicts that a stable Bose system (superfluid or normal)  will have a {\em fermionic} energy density in the unitarity limit, a fact can be tested by experiments. 

This work is supported by NASA GRANT-NAG8-1765  and NSF Grant DMR-0109255.

\noindent Figure caption: The upper plot determines the superfluid (SF) and 
normal (N) regions with densities eq.(\ref{ns}) and (eq.(\ref{nint})) depicted as 
solid and dashed curves in the lower figure.
The arrows in (N) indicate increasing degenerate (D) and Boltzmann (B) behavior. 
We use  $\mu(r) = \mu-M\omega^2 r^2/2$, 
 $A=1.3$, $B=1$, $D=1.2$, $\gamma=0.2$,  $\mu=32\hbar \omega$, $k_{B}T= 5\hbar \omega$, hence $\tilde{\gamma}=\gamma A=2.6$, $k_{B}T/\mu=0.16$, $k_{B}T/(\gamma A\mu)=0.65$.


\begin{thebibliography}{99}
\bibitem{Holland} M. Holland, et.al,  Phys. Rev. Lett. 87, 120406 (2001); M.L. Chiofalo et.al, 
Phys. Rev. Lett. 88, 090402 (2002)
\bibitem{Heiselberg} H. Heiselberg, Phys. Rev. A. 63, 043606 (2001).
\bibitem{HoMueller} T.L. Ho and E. Mueller, cond-mat/0306187.
\bibitem{Griffin} Y. Ohashi,  A. Griffin, Phys. Rev. Lett. 89, 130402 (2002). 
%\bibitem{Combescot}  Combescot
\bibitem{Duke}  K. M. O'Hara, et.al.,  {\em Science} {\bf 298}, 2179 (2002). M.E. Gehm, et.al. cond-mat/0212499.
\bibitem{ENS} T. Bourdel,  et.al, cond-mat/0303079. 
\bibitem{MIT}  4. S. Gupta et al, Science 300, 1723 (2003).
\bibitem{Jin1}  C. Regal and D. Jin, Phys. Rev. Lett. 90, 230404 (2003).
%\bibitem{Jin}  7. C. A. Regal, C. Ticknor, J. L. Bohn, D. S. Jin, Nature 424, 47 (2003).
%\bibitem{Rice}  K. E. Strecker, G. B. Partridge, R. G. Hulet, to appear in Phys. Rev. Lett.
%\bibitem{ENS2}  J. Cubizolles et al,  cond-mat/0308018.
\bibitem{LL}  E. Beth and G.E. Uhlenbeck, Physica, {\bf 4}, 915 (1937). See also p.232 in Landau and Lifshitz,  {\em Statistical Physics}, Addison Wesley 1974.
\bibitem{Fermif} $f_{3/2}(z)$$=$$\frac{4}{\pi^{1/2}}$$\int^{\infty}_{0} {\rm d} x x^2 
/(1+e^{x^2}/z)$$=$ $-\sum_{\ell=1}^{\infty}$ $(-z)^{\ell}/\ell^{3/2}$.
\bibitem{rhos} We write $K|\Psi|^2 = \rho^{}_{s}{\bf v}_{s}^{2}/2$, where ${\bf v}_{s}=\hbar\nabla \theta/(2M)$, and $\theta$ is the phase $\Psi$, $\Psi= |\Psi|^{i\theta}$.
\bibitem{second} Ch.14, I.M. Khalatnikov, {\em An Introduction to the Theory of Superfluidity} 
W.A. Benjamin, Inc. New York, 1965.
\bibitem{commentcal}  Since $\rho_{s}\propto (1-T/T_{c})$, to find $u_{2}^{}$ near $T^{}_{c}$, it is sufficient to evaluate 
$\sigma$ at $T_{c}$ and take $\rho_{n} = Mn$. Moreover, for Fermi systems at low temperatures, 
one can replace $(\partial \sigma/\partial T)_{P}$ by $(\partial \sigma/\partial T)_{n}$, and
$(\partial P/\partial T)_{\sigma}$ by $(\partial P/\partial T)_{n}$. 
\bibitem{1D} 1D Tonk gas has precisely the condition (infinite coupling constant) that 
motivates Universality Hypothesis. It has a fermionic energy density but not superfluidity. 
\end{thebibliography}
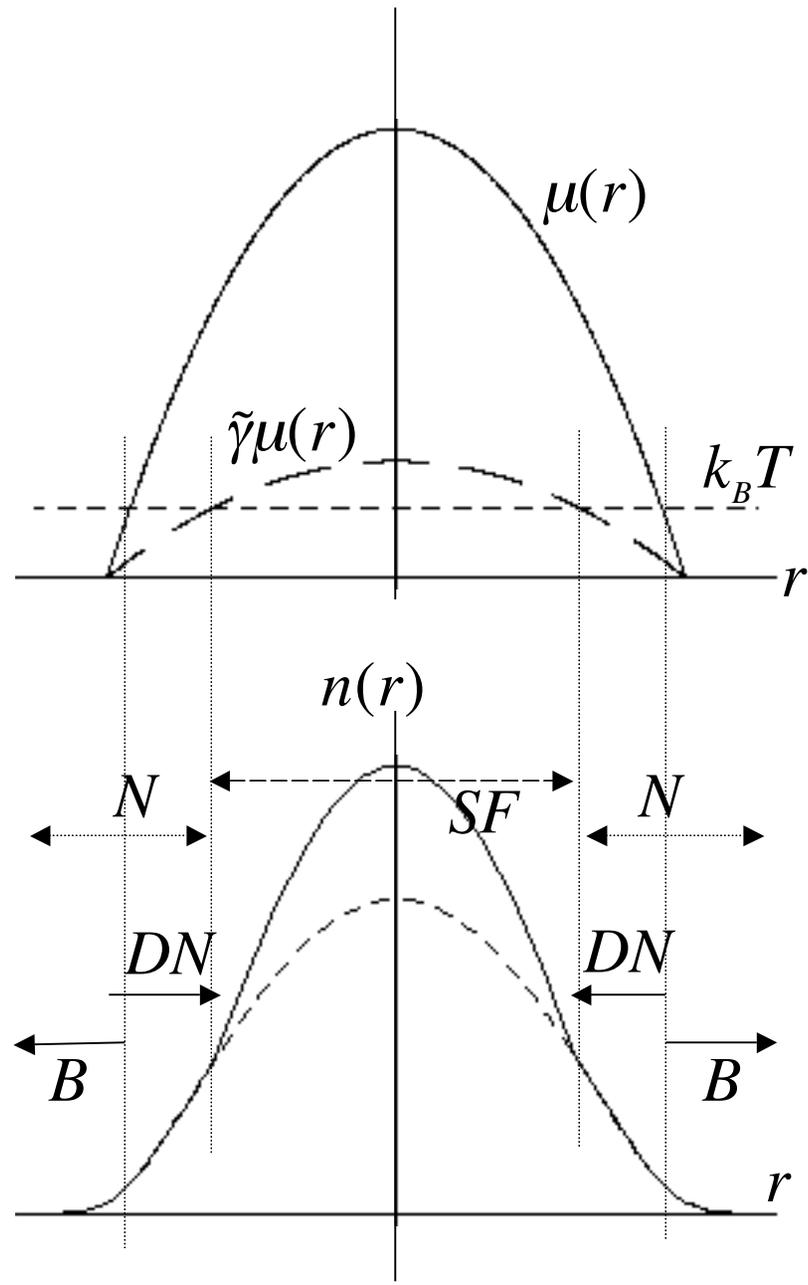
\end{document}